\magnification=1200 

\headline{\ifnum\pageno=1 \nopagenumbers 
\else \hss\number \pageno \fi} 
 
\overfullrule=0pt
\footline={\hfil}
\font\boldgreek=cmmib10
\textfont9=\boldgreek
\mathchardef\myphi="091E
\def\bfphi{{\fam=9 \myphi}\fam=1}
\def\lsim{\raise0.3ex\hbox{$<$\kern-0.75em\raise-1.1ex\hbox{$\sim$}}}
\def\gsim{\raise0.3ex\hbox{$>$\kern-0.75em\raise-1.1ex\hbox{$\sim$}}}
\baselineskip=10pt 
\vbox to 1,5truecm{}
\parskip=0.2truecm 
\centerline{\bf SIMPLE AMPLITUDES FOR $\bfphi^{\bf 3}$ FEYNMAN LADDER GRAPHS}
\medskip
 \centerline{\bf }\bigskip 

\bigskip \centerline{by}\smallskip 
\centerline{\bf R. Hong Tuan}  
\medskip
 \centerline{Laboratoire de Physique Th\'eorique et Hautes Energies
\footnote{*}{Laboratoire associ\'e au Centre National de la Recherche
Scientifique - URA D0063}}  \centerline{Universit\'e de Paris XI, b\^atiment 211, 91405
Orsay Cedex, France}  
\bigskip\bigskip
\baselineskip=20pt 
\noindent 
${\bf Abstract}$ \par 
Recently, we proposed a new approach for calculating Feynman graphs amplitude using the
Gaussian representation for propagators which was proven to be exact in the limit of
graphs having an infinite number of loops. Regge behaviour was also found in a
completely new way and the leading Regge trajectory calculated. Here we present symmetry
arguments justifying the simple form used for the polynomials in the Feynman parameters
$\bar{\alpha}_{\ell}$, where $\bar{\alpha}_{\ell}$ is the mean-value for these
parameters, appearing in the amplitude for the ladder graphs. (Taking mean-values is
equivalent to the Gaussian representation for propagators). \par

\vbox to 4 truecm{}

\noindent LPTHE Orsay 96/62 \par 
\noindent July 1996

\vfill\supereject
Quite a long time ago, ladder graphs$^1$ and the Bethe-Salpeter equations$^{2,3,4}$
were used to derive Regge behaviour and Regge trajectories in the context of scalar
$\phi^3$ field theory. Some years ago, a renewal of interest for these graphs was to
be noted$^{5,6}$. In the same period of time$^7$, we derived a rigorous derivation of
the Gaussian representation for propagators in superrenormalizable scalar field
theories which is valid in the infinite number of loops limit. This Gaussian
representation is equivalent to fixing the Feynman $\alpha$-parameter $\alpha_{\ell}$
of some propagator $\ell$ to a constant value $\bar{\alpha}_{\ell}$ which will be, in
fact, the mean-value for this parameter. \par

Then, once an overall scale for all $\alpha_{\ell}$'s has been separated and integrated
over one gets a particularly simple expression$^7$ for any Feynman graph amplitude $G$,
when $I$ the number of propagators and $L$ the number of loops tend to infinity, which
reads (the coupling $\gamma$ is taken equal to - 1) in the Euclidean region
$$F_G = (4 \pi )^{-dL/2} h_0 \left [ P_G \{\bar{\alpha}\} \right ]^{-d/2} \left [ Q_G (P,
\{\bar{\alpha}\}) + m^2 h_0 \right ]^{-(I-dL/2)} \Gamma (I - dL/2) h_0^{I-1}/(I-1)~!
\eqno(1)$$
\noindent where $h_0 = \sum\limits_{\ell} \bar{\alpha}_{\ell}$ is an arbitrary
constant and $h_0^{I-1}/(I-1)~!$ is the phase-space volume of the $\alpha_{\ell}$'s.
(This expression has been obtained by applying the mean-value theorem for the $I$
variables $\alpha_{\ell}$). $P_G(\{\alpha \})$ and $Q_G(P, \{\alpha\})$ are defined by
$$P_G(\{ \alpha \}) = \sum_{{\cal T}} \prod_{\ell \not\subset {\cal T}} \alpha_{\ell}
\eqno(2a)$$
$$Q_G(P, \{\alpha\}) = P_G^{-1}(\{ \alpha \}) \sum_{C} s_C \prod_{\ell \subset C}
\alpha_{\ell} \eqno(2b)$$
\noindent where the sum $\sum\limits_{{\cal T}}$ is the sum over all spanning trees of
$G$ (a spanning tree of $G$ is incident with every vertex of $G$) and $\sum\limits_C$
is the sum over all cuts $C$ of $G$ (a cut $C$ is the complement of a spanning tree
${\cal T}$ plus one propagator which cuts ${\cal T}$ into two disjoint parts). External
momenta $P_v$ are used to define
$$s_C = \left ( \sum_{v \in G_1} P_v \right )^2 = \left ( \sum_{v \in G_2} P_v \right )^2
\eqno(3a)$$
\noindent where $G_1$ and $G_2$ are two disjoint parts of $G$ (which are bound together
by the propagator cutting ${\cal T}$ used above to define ${\cal C}$). In a recent
letter$^8$ we derived the Regge behaviour for the sum of infinite ladders and the
corresponding leading Regge trajectory. But we needed the expressions of
$P_G(\{\bar{\alpha}\})$ and $Q_G(P, \{\bar{\alpha}\})$ for ladder graphs. In fact, we
displayed the very simple expressions
$$P_G(\{\bar{\alpha}\}) = (\bar{\alpha}_-)^L \exp (yL) \ f(y) \eqno(4a)$$
$$Q_G(P, \{\bar{\alpha}\}) = (t/2) \ L \ \bar{\alpha}_+ + s \ \bar{\alpha}_- \exp (- yL)
[f(y)]^{-1} \eqno(4b)$$
\noindent with
$$f(g) = {1 \over 2} y \left ( 1 + y^{-1} \right )^2 \eqno(5a)$$
$$y = \left ( 2 \bar{\alpha}_+/\bar{\alpha}_- \right )^{1/2} \eqno(5b)$$
\noindent $\bar{\alpha}_+$ being the mean-value of the $\alpha_{\ell}$'s for the
propagators parallel to the ladder and $\bar{\alpha}_-$ being the mean-value of the
$\alpha_{\ell}$'s for the central propagators. \par

It will be the purpose of the present letter to derive the expressions (4a) and (4b)
for $P_G(\{ \bar{\alpha}\})$ and $Q_G(P, \{ \bar{\alpha}\}$ as well as expressions
for some useful parameters. To understand how $P_G(\{\bar{\alpha}\})$ is obtained let
us look first at fig. 1a) where a ladder with $L$ loops is displayed. In order to
obtain a tree ${\cal T}$ from such a ladder we have to delete $L$ propagators whose
$\alpha_{\ell}$'s will be making one product in (2a). There are two different ways to
delete propagators, either in the center of the ladder bringing up a factor
$\bar{\alpha}_-$ or on the sides bringing up a factor $\bar{\alpha}_+$. If we
consider the removing of $p$ propagators in the center, then $(L - p)$ will be
removed on the sides. Each time we remove a propagator in the center we are fusing
two neighbour loops into what we call a ``cell'' which is built by propagators
surrounding it, making it a ``closed cell'. An open cell is built when one
propagator on the border of the closed cell is removed. \par

The three different topologies shown in fig. 1 are a) the case where only closed cells
are made removing center propagators, b) the case where one open cell is made removing
one end propagator c) the case where both end center
propagators are removed. Then, each remaining closed cell has
to be opened by removing one propagator on the sides. If the number of
side-propagators of a closed cell is $2 \ell_i$ (for a ``length'' $\ell_i$ for this
closed cell), then, there are $2 \ell_i$ ways of removing a side-propagator from a
closed cell to make an open cell. Once only open cells remain there is no way to
remove again propagators because we have already a tree ${\cal T}$. The three
topologies give rise to contributions to $P_G(\{ \bar{\alpha}\})$ that are respectively
$$P_G^{(a)}(\{ \bar{\alpha}\}) = \sum_{p=0}^{L-1} (\bar{\alpha}_-)^p
(\bar{\alpha}_+)^{L-p} \sum_{\ell_1 + \cdots + \ell_{L-p}=L} 2^{L-p} \ \ell_1 \cdots
\ell_{L-p} \eqno(6a)$$
$$P_G^{(b)}(\{ \bar{\alpha}\}) = 2 \sum_{p=1}^{L} (\bar{\alpha}_-)^p
(\alpha_+)^{L-p} \sum_{\ell_1 + \cdots + \ell_{L-p+1}=L} 2^{L-p} \ \ell_2 \cdots
\ell_{L-p+1} \eqno(6b)$$
$$P_G^{(c)}(\{ \bar{\alpha}\}) = \sum_{p=2}^{L} (\bar{\alpha}_-)^p
(\alpha_+)^{L-p} \sum_{\ell_1 + \cdots + \ell_{L-p+2}=L} 2^{L-p} \ \ell_2 \cdots
\ell_{L-p+1} \eqno(6c)$$
\noindent where $p$ center-propagators have been removed and $L - p$ side-propagators
have also been removed. Remark that for symmetry reasons given below we have taken the
same value $\bar{\alpha}_-$ for all mean-values of $\alpha_{\ell}$'s belonging to
center propagators and also the same value $\bar{\alpha}_+$ for all mean-values of
$\alpha_{\ell}$'s belonging to side propagators. We will see next that, indeed,
$P_G(\{\bar{\alpha}\})$ is a symmetric function of all
 $\bar{\alpha}_-$'s and of all $\bar{\alpha}_+$'s. The same will also be true for the
numerator of $Q_G(P, \{\bar{\alpha}\})$, $\sum\limits_C s_C \prod\limits_{\ell \subset
C} \alpha_{\ell}$, (see (2b)), where for the ladder, $s_C$ is either $t$, the
momentum-transfer square invariant or $s$, the large invariant when looking for Regge
behaviour (its logarithm is known in phenomenology, as the total ``rapidity''). \par

Now, let us look at the sums in (6), not making any assumption about the mean-values 
$$\sum_p 2^{L-p} \sum_{\{\ell_j\}} \ \ \sum_{\{i_+\} , \{i_-\}} \ \ \prod_{i_+ \in \{
i_+\}} \bar{\alpha}_{i_+} \prod_{i_- \in \{ i_-\}} \bar{\alpha}_i \prod_{\ell_k \in \{
\ell_k\}} \ell_k \eqno(7)$$
\noindent with the constraints $\sum\limits_i \bar{\alpha}_i = h_0$, $\sum\limits_k
\ell_k = L$, $\{i_+\}$ containing $L - p$ elements and $\{i_-\}$ containing $p$
elements (or indices). It is obvious that any two $\bar{\alpha}_{i_+}$'s with $i_+ \in
\{ i_+ \}$ can be interchanged leaving (7) unchanged. The same is true for any two
$\bar{\alpha}_{i_-}$'s with $i_- \in \{i_- \}$. The same is again true when
interchanging two $\bar{\alpha}_{i_+}$'s with $i_+ \notin \{i_+\}$ or two
$\bar{\alpha}_{i_-}$'s with $i_- \notin \{i_-\}$. The more delicate case arises
interchanging two $\bar{\alpha}_+$'s with one $i_+ \in \{i_+\}$ and the other $i_+
\notin \{i_+\}$ or two $\bar{\alpha}_-$'s with one $i_- \in \{ i_-\}$ and the other
$i_- \notin \{i_-\}$. \par

Let us take first the $(-)$ case. We can decompose $\{i_-\}$ in $\{i'_-\} \cup j_-$
with
$$\sum_{\{i_-\}} \prod_{i_- \in \{ i_-\}} \bar{\alpha}_{i_-} = \sum_{j_- \notin
\{i'_-\}} \ \ \sum_{\{i'_-\}} \bar{\alpha}_{j_-} \prod_{i'_- \in \{i'_-\}}
\bar{\alpha}_{i'_-} \ \ \ . \eqno(8)$$
\noindent So $\bar{\alpha}_{j_{1-}}$ with $j_{1-} \in \{ i_- \}$, $\notin \{ i'_- \}$
will be exchanged with $\bar{\alpha}_{j_{2^-}}$ with $j_{2^-} \notin \{i_-\}$, this for
a specific $\{i_-\}$. For another $\{i_-\}$ the roles of $\bar{\alpha}_{j_{1-}}$ and
$\bar{\alpha}_{j_{2^-}}$ might be reversed. Then, in the sum over all ensembles
$\{i_-\}$, we have an invariance under the interchange $\bar{\alpha}_{j_{1-}}
\leftrightarrow \bar{\alpha}_{j_{2^-}}$ (we remark that no $\{i'_-\}$ contains
$j_{1-}$ nor $j_{2-}$ and that $\{i_-\}$ can be either $\{i'_-\} \cup j_{1-}$, or
$\{i'_-\} \cup j_{2-}$, or $\{i'_-\} \cup j_{k-}$ with $j_{k-} \not= j_{1-}$,
$j_{2-}$). The reasoning concerning the exchange $j_{1^-} \leftrightarrow j_{2^-}$ would
be over if the factor $\prod\limits_{i_+ \in \{ i_+ \}} \bar{\alpha}_{i_+}
\prod\limits_{\ell_k \in \{\ell_k\}} \ell_k$ in (7), multiplying (8), was invariant under
this exchange. This is what we examine next. We displayed in fig. 2 the two
configurations discussed above for the exchange $\bar{\alpha}_{j_{1^-}} \leftrightarrow
\bar{\alpha}_{j_{2-}}$. \par

The total length of the cells enclosing $j_{1-}$ is $L_{k_1}$ and the total length of
the cells enclosing $j_2$ is $L_{k_2}$. Because there is only one $\bar{\alpha}_{i_+}$
factor for each cell, we see that going from a) to b) (i.e. making the exchange
$j_{1-} \leftrightarrow j_{2-}$) we have to suppress one $\bar{\alpha}_{i_+}$ in one of
the cells enclosing $j_{1-}$ in a) and put an additional one in one of the cells
enclosing $j_{2-}$ in b). This amounts to make an exchange $i_{1+} \leftrightarrow
i_{2+}$ or $\bar{\alpha}_{i_{1+}} \leftrightarrow \bar{\alpha}_{i_{2+}}$. So, if we have
symmetry under this exchange, we have symmetry under the exchange $\bar{\alpha}_{j_{1-}}
\leftrightarrow \bar{\alpha}_{j_{2-}}$ provided the last factor $\prod\limits_{\ell_k\in
\{\ell_k\}} \ell_k$ is also symmetric. And, in fact, going from a) to b) we have to make
an interchange in the order of $\ell_k$ factors under which $\prod\limits_{\ell_k \in \{
\ell_k\}} \ell_k$ is invariant and then, an exchange of the lengths $L_{k_1}
\leftrightarrow L_{k_2}$ under which $\sum\limits_{\{\ell_k\}} \prod\limits_{k \in
\{\ell_k\}} \ell_k$ is invariant because $L_{k_1}$ and $L_{k_2}$ are summed over (they
are ``dummy variables''). So, indeed, symmetry under interchange of the
$\bar{\alpha}_{i_+}$'s entails symmetry under interchange of the $\bar{\alpha}_{i_-}$'s.
Now, we will argue that all $\bar{\alpha}_{i_+}$'s should be given the same value. \par

We proved some years ago$^7$ that all $\bar{\alpha}_{\ell}$'s can be written
$\bar{\alpha}_{\ell} = C_{\ell} (h_0/I)$ where $C_{\ell}$ is some constant not
depending on $I$. The basic reason for this is given by the simple factoring of the
total phase-space $h_0^{I-1}/(I - 1)~!$ into $I$ factors which gives for each
$\alpha_{\ell}$ a phase-space $\simeq e h_0/I$. \par

The equations$^{7,8}$ which determine $C_{\ell}$ make it depend on $t$ and $s$ (or in
general any invariant $s_C$) through $Q_G(P, \{\bar{\alpha}\})$ which is a ratio of
two polynomials of degree $L + 1$ (for $\sum\limits_{C} s_C \prod\limits_{\ell \subset
C} \alpha_{\ell}$) and degree $L$ (for $P_G(\{ \bar{\alpha}\})$). Therefore $Q_G(P, \{
\bar{\alpha}\})$ is homogeneous to only one power of $\bar{\alpha}_{\ell}$ and behaves
a priori like $O(h_0/I)$. However, in the special case of ladders the number of ways
we can cut across any ${\cal T}$ in order to get $s_C = t$ is in fact $L$ (each open cell
of length $\ell_i$ can be cut $\ell_i$ times) which is proportional to $I$. (And,
we have to multiply by 1/2 in order to avoid the double-counting of cuts). So we have
$Q_G(P, \{\bar{\alpha}\})$ proportional to $t \ C_+(L/2)(h_0/I) \sim C_+t$. (The second
term is negligible because the factor multiplying $s$ is proportional to $(h_0/I)/\exp
(C^{st}.I)$, the exponential coming from the exponential number of spanning trees ${\cal
T}$ in $G$ as $I \to \infty$. This is visible in the final expression (4b) for $Q_G(P,
\{\bar{\alpha}\})$. In the Regge limit where $s \to \infty$, the saddle point one finds
when summing over all $L$ gives $s \exp (-yL) = C^{st}$ or $L \simeq \ell n s$ so that
even in this case the $\bar{\alpha}_-$ factor $\sim (h_0/I)$ kills the term proportional
to $s$). So, we conclude that the $C_{\ell}$'s, in the case of ladders {\it only depend
on the invariant $t$}. Now, assuming that $\bar{\alpha}_{i_+}$ varies along the ladder
means that the corresponding constant, $C_+$, should also depend on $x_i = L_i/L$ if
$L_i$ is the number of propagators along one side of the ladder separating the
propagator $i_+$ from, for instance, an external line on the left of the ladder.  \par

Now, let us isolate a sub-ladder $G_1$ (of the ladder $G$) with length $L_1$ (with $L_1
\to \infty$). We note that for $G_1$, the invariant $t$ is the
same as for $G$, because $t$ is conserved along the ladder $G$. So, $C_+$, along $G_1$
only depends on $x_{i_1} = L_{i_1}/L_1$, if $L_{i_1}/L_1$ is the distance of an $i_+$
propagator to the left end of $G_1$. And the function $C_+(x_{i_1}, t)$ should be the
same for $G$ and $G_1$, the equations determining it being the same. That is, we should
have 
$$C_+(x_i, t) = C_+ (x_{i_1}, t) \eqno(9)$$
\noindent for any sub-ladder $G_1$, which entails that $C_+$ should not depend on
$x_i$, i.e. $\bar{\alpha}_{i_+}$ should be the same along $G$, and therefore
$\bar{\alpha}_{i_-}$ too according to the first part of our argument. \par

Of course, writing (9), we implied that the amplitude for $G_1$ factorized from the
rest of $G$. However, we proved$^{9}$ that, precisely, the weighted sum of all spanning
trees ${\cal T}$ of a graph $G$ containing an infinite number of loops could be
factorized into independent factors for a connected sub-graph containing also an
infinite number of loops and its complement graph on $G$. $G$ and $G_1$ having an
infinite number of loops, and $G_1$ being connected, the factorization property holds in
the present case. (In the more restraint domain of multiperipheral dynamics$^3$ this
factorization is also implied by the well-known short range correlation property along
ladders). \par

A straightforward way of seeing the factorization appear in the case of ladders is to
consider the amplitude for $G_1$, $A_{G_1}(t, L)$, and to couple it through integration
over the momenta of its external lines to its neighbour ladders $G_0$ and $G_2$. Because
$A_{G_1}(t, L)$ only depends on $t$ and $L$, integrating over the momenta in the loops
at the ends of $G_1$, we see an {\it exact} factorization appear ($g(t)$ is the coupling
between two adjacent ladders) $$A_G(t, L) = A_{G_0}(t, L_0) \ g(t) \ A_{G_1}(t, L_1) \
g(t) \ A_{G_2}(t, L_2) \eqno(10)$$ \noindent which justify its use in deriving (9). This
happens because {\it the dependence over the other invariant ($s_1$ for $G_1$) only
appears through the evaluation of the sum over all $L_1$'s}, giving a saddle-point$^8$
of $L_1 \simeq \ell n s_1$ when $s_1$ is allowed to go to infinity. \par 

For $Q_G(P, \{\bar{\alpha}\})$, the factor $\sum\limits_{C} s_C \prod\limits_{\ell
\subset C} s_C$ in (2b) can be decomposed like $P_G(\{\bar{\alpha}\})$ into three
distinct contributions corresponding to the three topologies of fig. 1. These are
$$\widetilde{Q}_G^{(a)}(P, \{\bar{\alpha}\}) = s \ \bar{\alpha}_-^{L+1} + (- 2
\bar{\alpha}_- m^2 + (t/2) L \bar{\alpha}_+) P_G^{(a)} (\{\bar{\alpha}\}) \eqno(11a)$$
$$\widetilde{Q}_G^{(b)}(P, \{\bar{\alpha}\}) = - \bar{\alpha}_- m^2 P_G^{(b)} (\{
\bar{\alpha}\}) + \sum_{\ell_1 = 1}^{L-1} \left [ - 2 \ell_1 m^2 + (t/2) (L - \ell_1)
\right ] \bar{\alpha}_+ \ P_{G, \ell_1}^{(b)}(\{ \bar{\alpha} \}) \eqno(11b)$$
$$\widetilde{Q}_G^{(c)}(P, \{\bar{\alpha}\}) = \sum_{\ell_1 = 1, \ell_{L-p+2=1}}^{L-2}
\left [ - 2 \left ( \ell_1 + \ell_{L-p+2} \right ) m^2 + (t/2) \left ( L - \ell_1 -
\ell_{L-p+2} \right ) \right ] \cdot$$  $$\bar{\alpha}_+ \ P_{G, \ell_1, \ell_{L-p+2}}
(\{ \bar{\alpha} \}) \eqno(11c)$$
\noindent where $P_{G, \ell_1}^{(b)}(\{ \bar{\alpha}\})$ refers to
$P_G^{(b)}(\{\bar{\alpha}\})$ with a fixed $\ell_1$ and $P_{G,\ell_1,
\ell_{L-p+2}}^{(c)}(\{\bar{\alpha}\})$ refers to $P_G^{(c)}(\{ \bar{\alpha}\})$ with
fixed $\ell_1$ and $\ell_{L-p+2}$. Summing over $\ell_1$, or over $\ell_1$ and
$\ell_{L-p+2}$, these quantities have finite mean-values that we can neglect
relatively to $L$. We then get
$$Q_G(P, \{ \bar{\alpha}\}) = (t/2) \ L \ \bar{\alpha}_+ + s \ \bar{\alpha}_-^{L+1} \left
[ P_G(\{\bar{\alpha}\}) \right ]^{-1} \ \ \ . \eqno(12)$$
\noindent It is clear that all the reasoning done specifically for
$P_G(\{\bar{\alpha}\})$ deriving the symmetry properties $\bar{\alpha}_{j_{1-}}
\leftrightarrow \bar{\alpha}_{j_{2-}}$ if there is a $\bar{\alpha}_{i_{1+}}
\leftrightarrow \bar{\alpha}_{i_{2+}}$ symmetry can be redone for $Q_G(\ell ,
\{\bar{\alpha}\})$, $\prod\limits_{j_- \in G} \bar{\alpha}_{j-}$ replacing
$(\bar{\alpha}_-)^{L+1}$ and $\sum\limits_{j_+ \notin \{i_+\}} \bar{\alpha}_{i_+}$
replacing $L \bar{\alpha}_+$. Our next step will be to perform the sums in (6) in
order to obtain a compact expression for $P_G(\{\bar{\alpha}\})$. \par

We use the formulae 
$$\int_{x_i \geq 0} \prod_{i=1}^n dx_i \ x_i^{p_i-1} \ \delta \left ( S - \sum_{i=1}^n
x_i \right ) = {\partial \over \partial S} I(S) \eqno(13a)$$
\noindent and$^{10}$
$$I(S) = \int_{x_i \geq 0} \prod_{i=1}^n dx_i \ x_i^{p_i-1} \theta \left ( S -
\sum_{i=1}^n x_i \right ) = S^{\sum\limits_{i=1}^n p_i} \prod_{i=1}^n \Gamma
(p_i)/\Gamma \left ( 1 + \sum_{i=1}^n p_i \right ) \ . \eqno(13b)$$
\noindent Replacing $S$ by $L$, $p_i$ by 2 and $n$ by $L - p$ we get 
$$\eqalignno{
P_G^{(a)}(\{ \bar{\alpha}\}) &= \sum_{p=0}^{L-1} (\bar{\alpha}_-)^p
(\bar{\alpha}_+)^{L-p} \ 2^{L-p} \ L^{2(L-p)-1}/[(L-p) - 1]~! \cr
& \simeq (\bar{\alpha}_-)^L (2 \bar{\alpha}_+/\bar{\alpha}_-)^{1/2} {\rm sh} \  \left [
(2 \bar{\alpha}_+/\alpha_-)L \right ] &(14a) \cr
}$$
$$P_G^{(b)}(\{ \bar{\alpha}\}) = 2 (\bar{\alpha}_-)^L \left \{ {\rm ch} \ \left [ (2
\bar{\alpha}_+/\bar{\alpha}_-)^{1/2} L \right ] - 1 \right \} \eqno(14b)$$
$$P_G^{(c)}(\{ \bar{\alpha}\}) = (\bar{\alpha}_-)^L (2
\bar{\alpha}_+/\bar{\alpha}_-)^{-1/2} \ {\rm sh} \ \left [ (2
\bar{\alpha}_+/\bar{\alpha}_-)^{1/2} L \right ]  \eqno(14c)$$
\noindent and therefore obtain the expressions (4) for $P_G(\{ \bar{\alpha}\})$ and
$Q_G(P, \{ \bar{\alpha}\})$. In the calculation of $\bar{\alpha}_{i_-}$ or
$\bar{\alpha}_{i_+}$$^{7,8}$ we decompose $P_G(\{ \bar{\alpha}\})$ into~:
$$P_G(\{ \bar{\alpha}\}) = a_i + b_i \bar{\alpha}_i = b_i (a_i/b_i + \bar{\alpha}_i)
\eqno(15)$$
\noindent where $i$ is either an $i_+$ or an $i_-$ propagator. \par

The formulae (13) then allow us to calculate $\mu_+ = a_+/b_+$ and $\mu_- = a_-/b_-$.
We have for $b_+ \bar{\alpha}_{i_+}$ a cell which is cut at $i_+$ and for $P_G(\{
\bar{\alpha}\})$ the same cell can be cut $2\ell$ times if the cell has length $\ell$.
Then
$$<1/(2\ell )> = b_+ \bar{\alpha}_{i_+}/P_G(\bar{\alpha}) = \bar{\alpha}_+/(\mu_+ +
\alpha_+) \eqno(16)$$
\noindent where $<1/(2 \ell )>$ is the mean-value of $1/(2 \ell )$ in the sum defining
$P_G(\{\bar{\alpha}\})$. Then, taking the mean-value, one has to replace
$\sum\limits_{i=1}^n p_i$ by $\left ( \sum\limits_{i=1}^n p_i \right ) - 1$, and
multiply by 1/2, which gives 
$$<1/(2 \ell )> = 1/2 \left ( 2 \bar{\alpha}_+/\alpha_- \right )^{1/2} = y/2
\eqno(17a)$$
$$\mu_+/\bar{\alpha}_+ = 2/y - 1 \ \ \ . \eqno(17b)$$
\noindent For the $(-)$ case, for $a_-$ we have the end propagator $i_-$ of a cell of
length $\ell_-$ which is not removed, while for $b_- \bar{\alpha}_{i_-}$ $\ (\ell_- - 1)$
propagators $i_-$ can be removed from the cell. This gives
$$<1/\ell > = \bar{a}_-/P_G(\{\bar{\alpha}\}) = \mu_-/(\mu_- + \bar{\alpha}_-) \ \ \ .
\eqno(18)$$
\noindent This amount again to replace $\sum\limits_{i=1}^n p_i$ by $\left (
\sum\limits_{i=1}^n p_1 \right ) - 1$ (without dividing by two), so we have
$$<1/\ell > = \mu_-/\left ( \mu_- + \bar{\alpha}_- \right ) = y \eqno(19a)$$
\noindent or
$$\mu_-/\bar{\alpha}_- = 1/(y^{-1} - 1) \ \ \ . \eqno(19b)$$
\noindent These expressions are crucial to define all parameters appearing in the
equations$^{7,8}$ defining $\bar{\alpha}_+$ and $\bar{\alpha}_-$.

\vfill \supereject \centerline{\bf References} \bigskip
 \item{1 -} J. C. Polkinghorne, J. Math. Phys. {\bf 4}, 503 (1962). 
 
\item{2 -} B. W. Lee and R. F. Sawyer, Phys. Rev. {\bf 127}, 2266 (1962). 

\item{3 -} L. Bertocchi, S. Fubini and M. Tonin, Nuovo Cimento {\bf 25}, 626 (1962)~;
D. Amati, A. Stanghellini and S. Fubini, Nuovo Cimento {\bf 26}, 6 (1962). 

\item{4 -} N. Nakanishi, Phys. Rev. {\bf B135}, 1430 (1964). 

\item{5 -} N. I. Ussyukina and A.I. Davydychev, Phys. Lett. {\bf B305}, 136 (1993). 

\item{6 -} R. Gastmanns, W. Troost and T. T. Wu, Nuclear Physics {\bf B365}, 404 (1994).

\item{7 -} R. Hong Tuan, Phys. Lett. {\bf 286}, 315 (1992).

\item{8 -} R. Hong Tuan, Regge behaviour and Regge trajectory for ladder graphs in
scalar $\phi^3$ field theory, Orsay preprint LPTHE 96/46.

\item{9 -} R. Hong Tuan, Factorization of Spanning Trees on Feynman Graphs, preprint
LPTHE Orsay 92/59.

\item{10 -} I. S. Gradshteyn and I. M. Ryzhik, Table of Integrals, Series and Products, 
Academic Press (New York) p. 620.
\vfill \supereject
\centerline{\bf Figure Captions} \bigskip
{\parindent=2 truecm
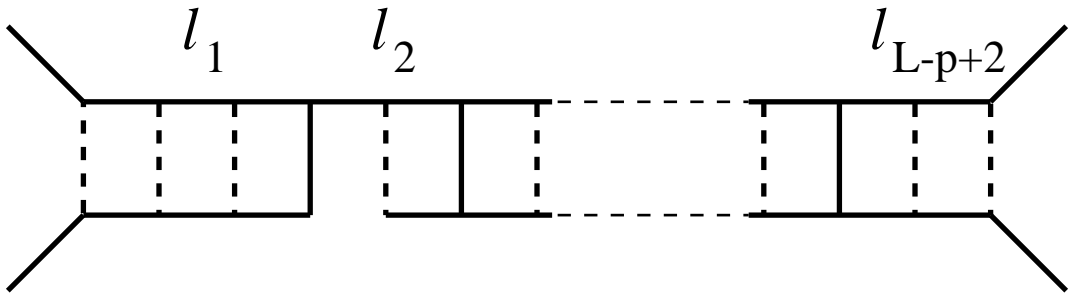
\item{\bf Fig. 1} We display the three topologies for spanning trees ${\cal T}$ in the
ladder graphs. $L - p$ counts the number of open cells of length $\ell_i$. Thick dashed
lines indicate removed $-$ propagators. Gaps along the ladder indicate
removed $+$ propagators. \par \vskip 3 truemm

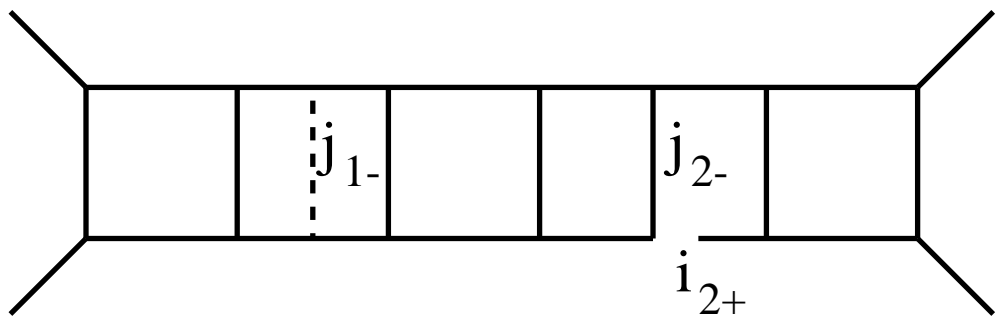
\item{\bf Fig. 2} A propagator $j_{1-}$ is exchanged with a propagator $j_{2-}$ (one
being removed and the other not) going from a) to b). A propagator $i_{1+}$ is also
exchanged with a propagator $i_{2+}$ to keep only one + removed propagator per cell.
The total length of the cells enclosing $j_{1-}$ in a) and $j_{2-}$ in b) are
respectively $L_{k_1}$ and $L_{k_2}$. \par \vskip 3 truemm

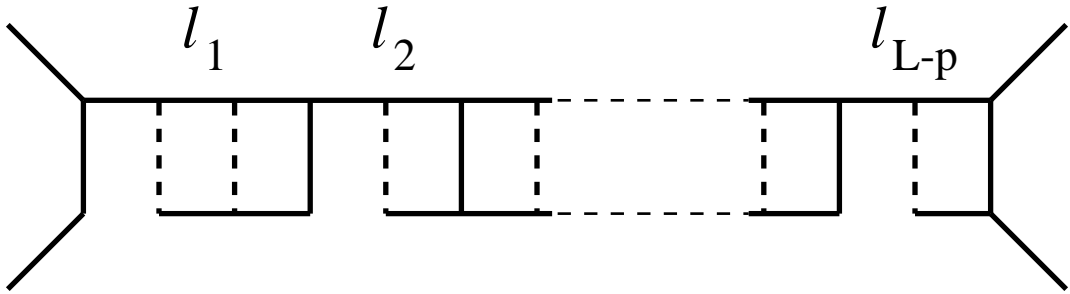
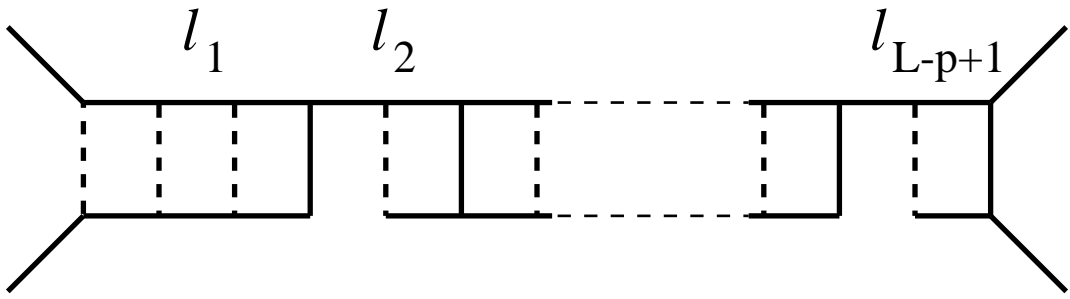
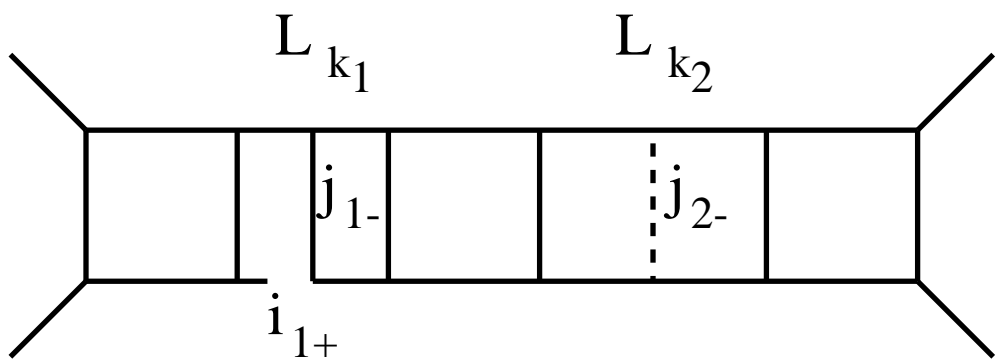
\bye